\documentstyle[11pt,mypasp,twoside,color,colordvi]{article}
\def\edcomment#1{\iffalse\marginpar{\raggedright\sl#1\/}\else\relax\fi}
\reversemarginpar

\begin{document}
\title{The Grandeur of Massive Star Formation --- \\ Revealed with ISAAC}
 \author{Hendrik Linz and Bringfried Stecklum}
\affil{
TLS Tautenburg, Sternwarte 5, D--07778
Tautenburg, Germany}
\author{Thomas Henning}
\affil{
MPIA Heidelberg, K\"onigstuhl 17, D--69117
Heidelberg, Germany}
\author{Peter Hofner}
\affil{New Mexico Tech, 801 Leroy Place, Socorro, N.M. 87801, U.S.A.}
\begin{abstract}

We present selected results from our ongoing investigation of high--mass
star--forming regions which are based on infrared observations with ESO's
ISAAC camera at the 8.2-m ANTU VLT telescope. Although these young stellar
objects comprise a high degree of complexity, our data enable us to disentangle
these crowded regions. By means of broad-- and narrow--band imaging between 1--5
micron we performed a thorough characterisation of the embedded population.
Special emphasis was put on the importance of an accurate astrometry which 
has a major impact on the interpretation of the data.
In the case of G9.62+0.19, we clarified the true nature of the infrared
emission in the immediate vicinity of an hot molecular core (HMC). We unveil 
the counterpart of this HMC in the thermal infrared. For GGD27 IRS2,
we found thermal emission at 3.8 and 4.7 micron caused by a deeply embedded
object that is powering a large radio jet and has counterparts in our
mid--infrared and VLA 7-mm data.
The presented results mark a further step on the way to disclose the mechanisms of
massive star formation. They demonstrate the value of sensitive infrared
imaging with the current generation of IR cameras on 8-m--class telescopes.

\end{abstract}

\section{Introduction}

The observations with the IR camera ISAAC (Moorwood 1997) were carried out in the 
first half of 2001 in service mode which was chosen to ensure that the ambient 
conditions meet our specifications, in particular the seeing demand. The narrow--band 
imaging in the 2.09 $\mu$m filter was performed in polarimetric mode, i.e., with the 
Wollaston prism and the slit mask in the optical train. The dithering positions 
were chosen to achieve the full coverage of the field--of--view (FOV). Furthermore, 
these images served as continuum which had to be subtracted from the Br$\,\gamma$ and 
H$_2$(1--0)S1 narrow--band frames. For the data reduction we developed a general pipeline 
applicable to imaging data obtained by ISAAC. The pipeline is written in the high--level 
IDL language, offering automated reduction together with the opportunity of flexibility. 
The aims of the measurements were as follows: 
\begin{itemize}
   \item Images in continuum filters (2.09 $\mu$m, L$'$, nb\_M) to study the distribution of stars 
         from the most massive to solar--type ones.
   \item Narrow--band imaging in Br$\,\gamma$ (2.166 $\mu$m) and Br$\,\alpha$ 
         (4.07 $\mu$m) to deduce the amount and distribution of dust extinction.
   \item Images in the H$_2$(1--0)S1 line (2.122 $\mu$m) to search for shocks caused by 
         winds and outflows from young stellar objects (YSOs).

\end{itemize}

Here, we show some results for two selected regions of high--mass star formation. 
Together with recent interferometric radio data, they demonstrate that high--resolution 
observations are required to overcome the confusion arising from the complexity of 
massive star formation. 

\section{Selected Results}

\subsection{G9.62+0.19 -- A hot molecular core puzzles the IR observers}

G9.62+0.19 is a complex of massive star formation regions (d$\, \sim \,$5.7 kpc). It
comprises several YSOs in different evolutionary stages. An age gradient spans from 
western (older) to eastern (younger) regions (Hofner et al. 1996). The site harbours an 
hot molecular core (HMC), i.e., a very dense, warm and compact condensation of molecular gas 
(Cesaroni et al. 1994). NIR emission was found at the location of the HMC (Testi et al. 
1998), in contradiction of HMC standard 
models (Osorio et al. 1999). The superior sensitivity and resolution of the new ISAAC data 
(VLT archive data for the filters J, H and Ks as well as our thermal infrared and NIR 
narrow--band data) lead to new conclusions: \\
{\bf High resolution} (seeing 0\farcs4 -- 0\farcs6){\bf :} An intriguing substructure of the HMC 
region F has been revealed (decomposition into 3 IR objects, see Fig. 1). \\
{\bf Highly accurate astrometry:} The peak positions of the HMC 
radio emission and of the strongest NIR peak in component F do not coincide (Fig. 1). 
A foreground NIR star (F2) at the HMC position fades at longer wavelengths. However, 
another source becomes apparent nearby (F4 in Fig. 1) which seems to be intrinsically 
associated with the HMC. \\
{\bf Working hypothesis:} The finding of a pole--on molecular outflow probably driven by 
the HMC (Hofner et al. 2001) explains at least qualitatively why we can see at all 
thermal IR emission from the HMC -- we benefit from the outflow's clearing effect. \\
{\bf Still discussions:} Contrary to De Buizer et al. (2003), we find that also the 
compact 10-micron source found in the HMC region by several authors corresponds to 
the HMC. It is identical with object F4 (see Fig. 2). \\
Furthermore, there are localised regions of Br$\,\gamma$ and H$_2$ 
emission as well as of reflected NIR continuum within G9.62+0.19 (Fig. 3) which 
demonstrates the complex structure of this region also on a larger scale. \\
The results of a comprehensive IR study about this star--forming region and its HMC will be given 
in Linz et al. (2003). 

\newpage


\includegraphics{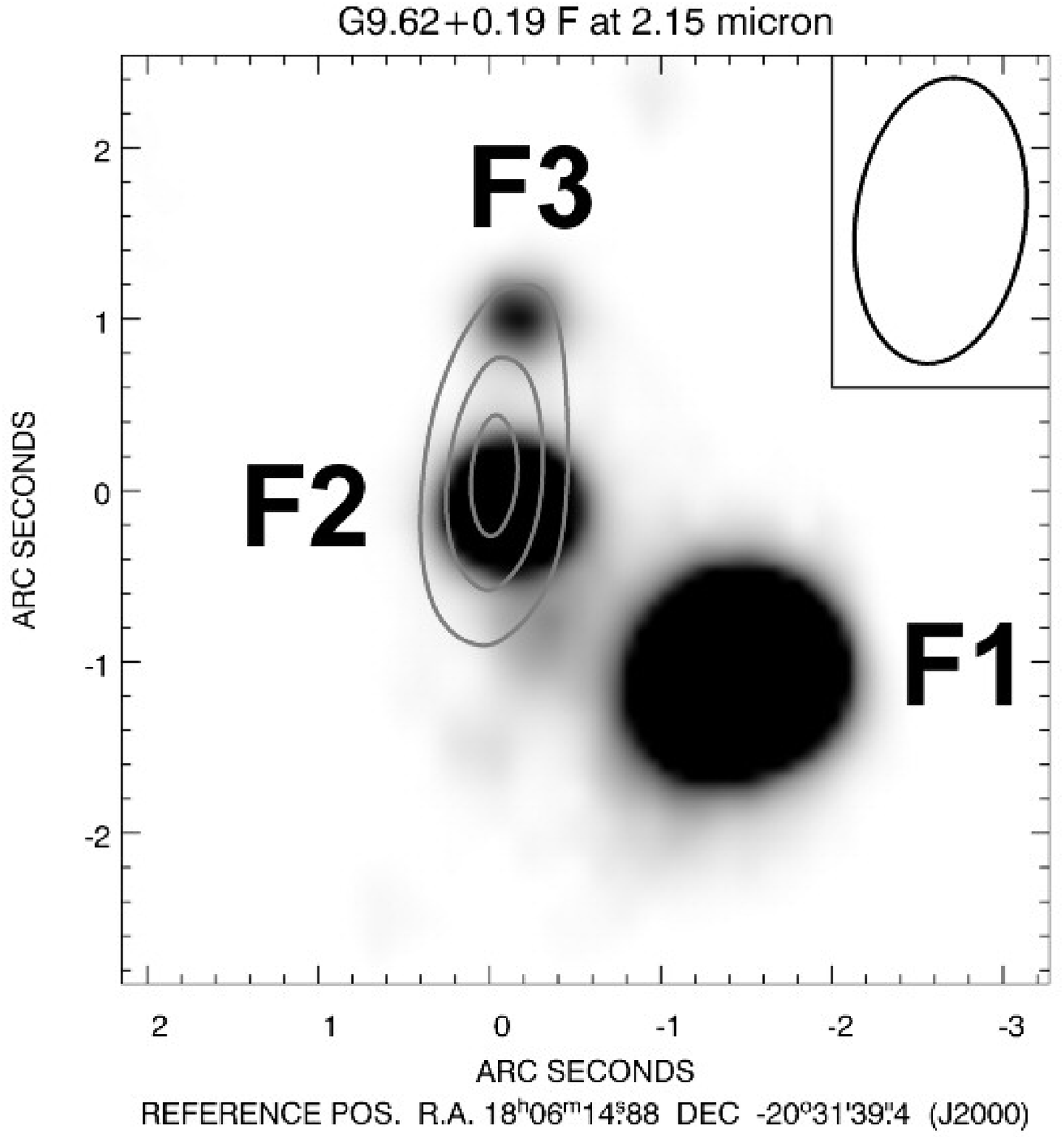}
\includegraphics{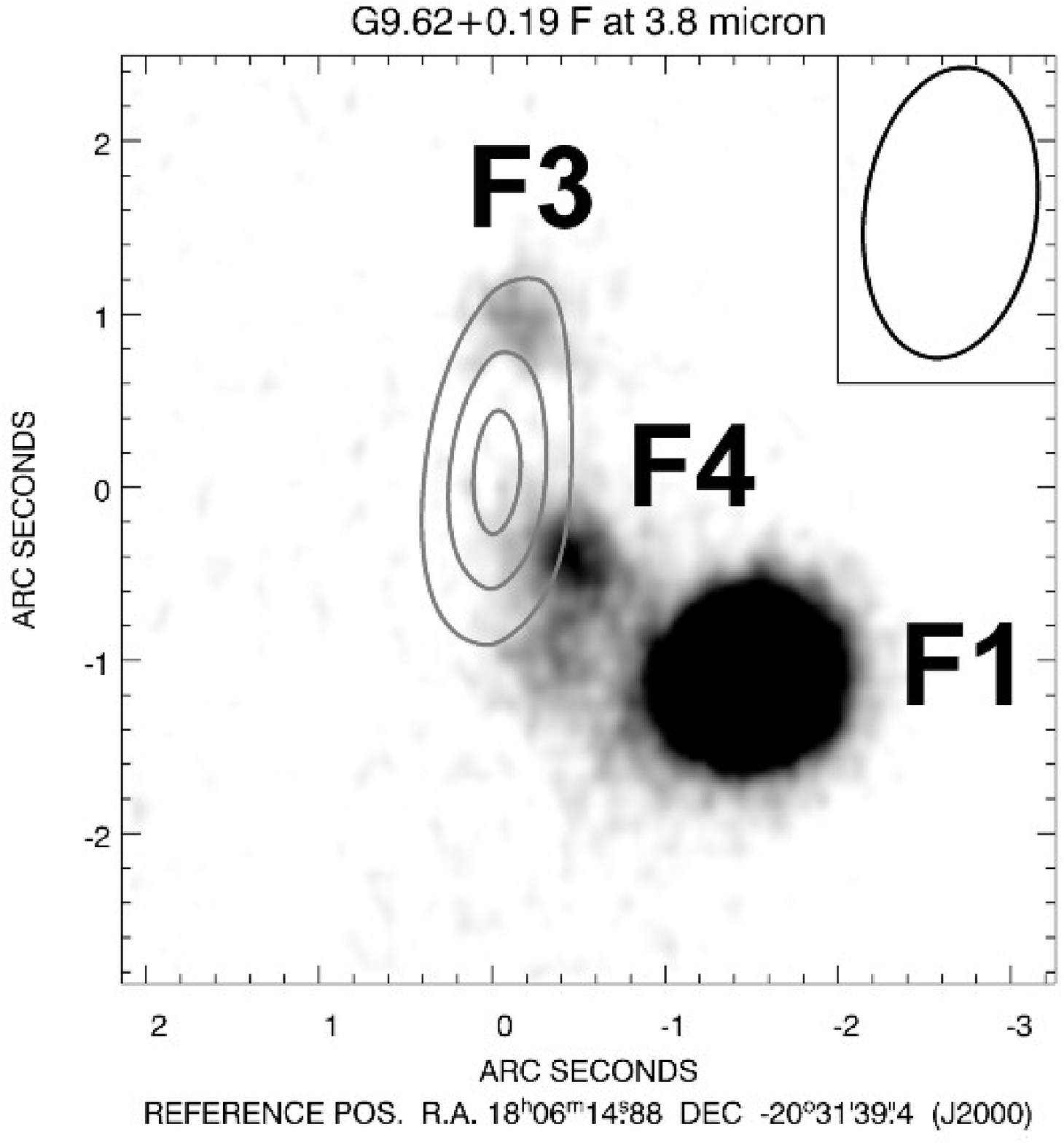}
\includegraphics{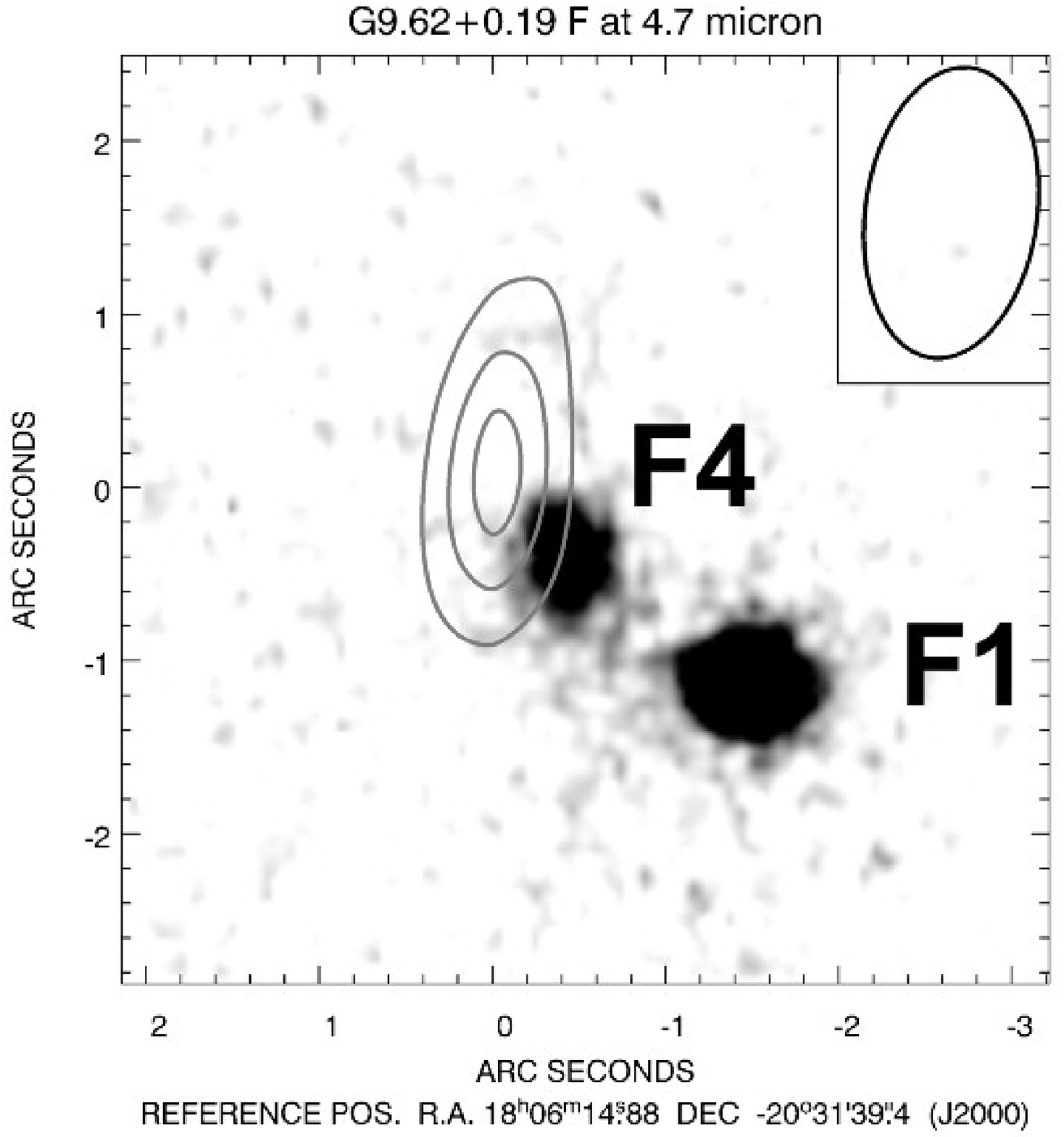}
\vspace*{5.0cm}
\begin{center}
\parbox{13cm}{\small \bf Figure 1. \rm \,\, Immediate vicinity around the the Hot Core region F in G9.62+0.19. 
From left to right: Ks band, L$'$ band, and narrow M band. The overlaid contours 
mark the peak position of the NH$_3$(5,5) emission of the hot molecular core 
(Hofner et al. 1994). The Ks band object F2 at this peak position turns out 
to be a foreground star that completely fades at longer wavelengths. But nearby 
we find strongly increasing emission (F4). We presume that this emission is 
associated with the Hot Core.}
\end{center}
\vspace*{1cm}
\includegraphics{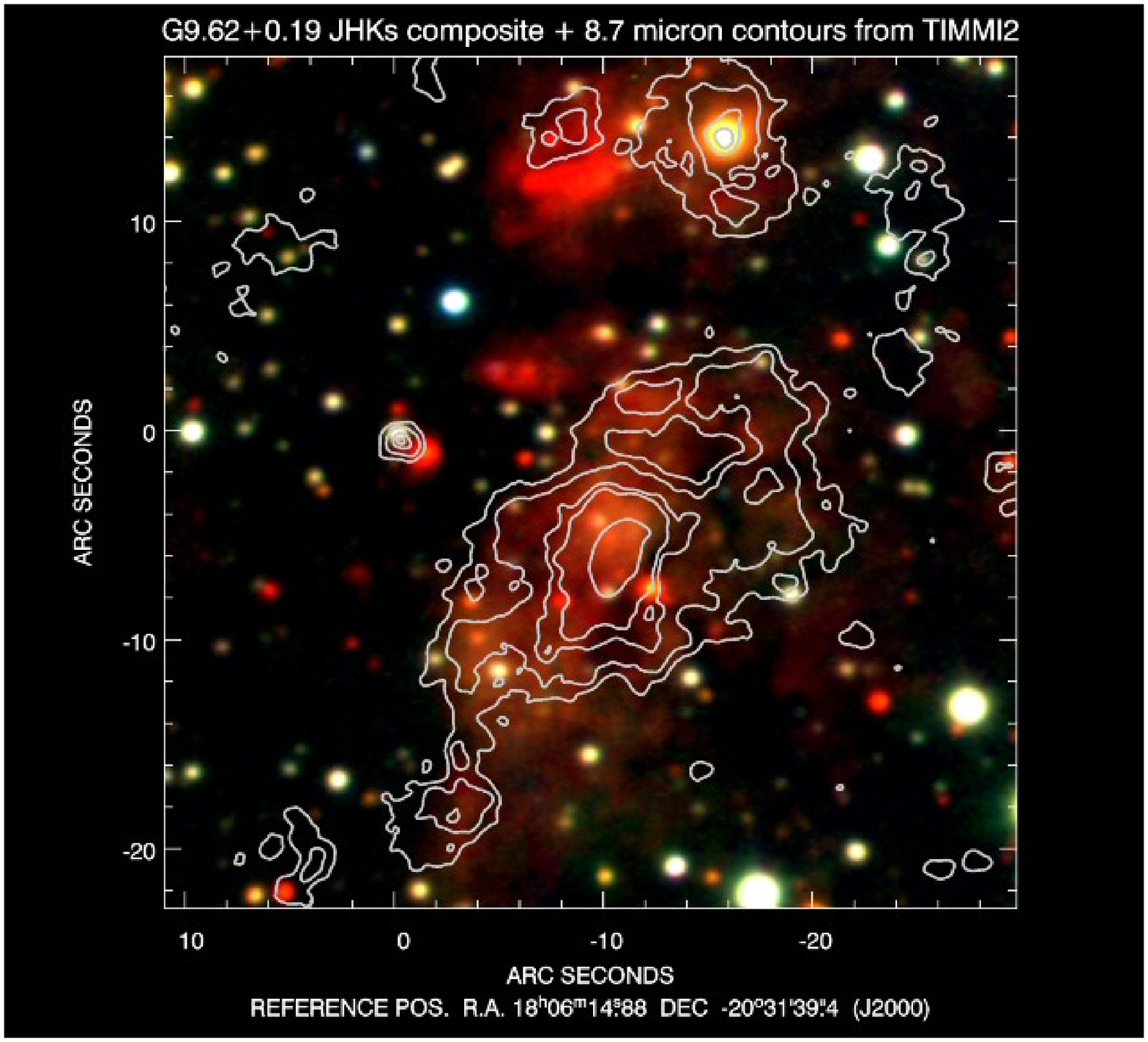}
\vspace*{10.0cm}
\begin{center}
\parbox{13cm}{\small \bf Figure 2. \rm \,\, Colour composite 
(\bf\textColor{1.0 1.0 0.0 0.1} J  band at 1.25 $\mu$m,
 \textColor{1.0 0.0 1.0 0.1} H  band at 1.65 $\mu$m, 
 \textColor{0.0 1.0 1.0 0.1} Ks band at 2.15 $\mu$m\rm\textColor{1.0 1.0 1.0 1.0}). 
 Overlaid are the 8.7 $\mu$m contours
derived from TIMMI2 observations we performed in July 2003 at ESO's 3.6-m telescope
in Chile. According to our astrometry, the compact mid--infrared object near the reference
position does not coincide with the (red) NIR source F1 (cf. Fig. 1), but probably traces 
the newly discovered object F4, i.e., the counterpart of the HMC. }
\end{center}

\newpage

\includegraphics{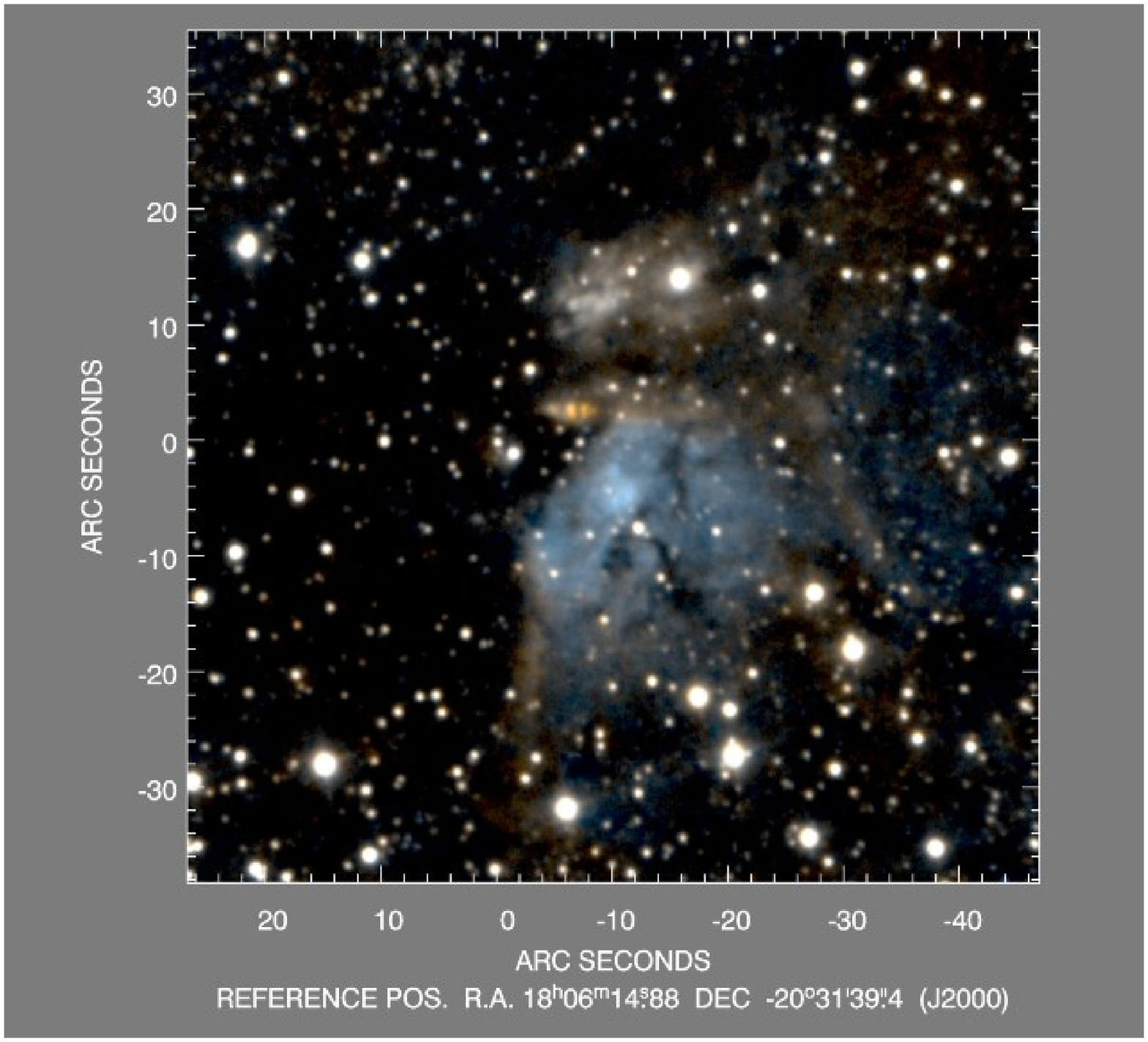}
\vspace*{10cm}
\begin{center}
\parbox{13cm}{\small \bf Figure 3. \rm \, \, Entire view on the star--forming complex G9.62+0.19, colour--coded 
with the H$_2$ emission (2.12 $\mu$m) in red and the Br$\,\gamma$ emission (2.17 $\mu$m) 
in blue. Beside the dominating (and expected) Br$\,\gamma$ emission arising from the 
H{\sc II} region (component B), this composite reveales the presence of a quite compact 
H$_2$ emission feature roughly in the centre of the image. The extended emission in the 
north (in white) is mainly reflected continuum emission.}
\end{center}

\subsection{GGD~27 -- A splendiferous IR reflection nebula with a 
            pumping heart}

This region, which contains a deeply embedded IRAS source in its centre, is 
associated with the Herbig--Haro objects HH80/81 and HH80-North (d$\,\sim\,$1.7 kpc). 
An IR reflection nebula surrounds the more enshrouded inner region. The central object 
was finally detected in the mid-IR (Stecklum et al. 1997). Mart\'{\i} et al. (1993) 
revealed a large and well--collimated thermal radio jet emanating from the central 
source. This jet--driving source is a prime candidate for the search of an accretion 
disk around a newly formed massive star. \\
Our new ISAAC NIR narrow-band and thermal infrared data show the nebula and its 
power source in unprecedented detail: The reflection nebula exhibits a bipolar
morphology (Fig. 4), the entire structure is roughly aligned along the NE--SW 
direction of the jet. A large fraction of the extended emission consists of
scattered light. Localised spots of H$_2$ emission (presumably due to shocks) exist. 
With the VLA, we found a 7-mm emission peak, indicative of thermal dust emission, 
that is coincident with the embedded jet--driving source and offset from NIR 
features (Figs. 4 and 5). 
In the thermal infrared, the driving source finally becomes visible (Fig. 5). With our new 
ISAAC data we could detect it for the first time at a wavelength as short as 
3.8 $\mu$m.

\newpage


\includegraphics{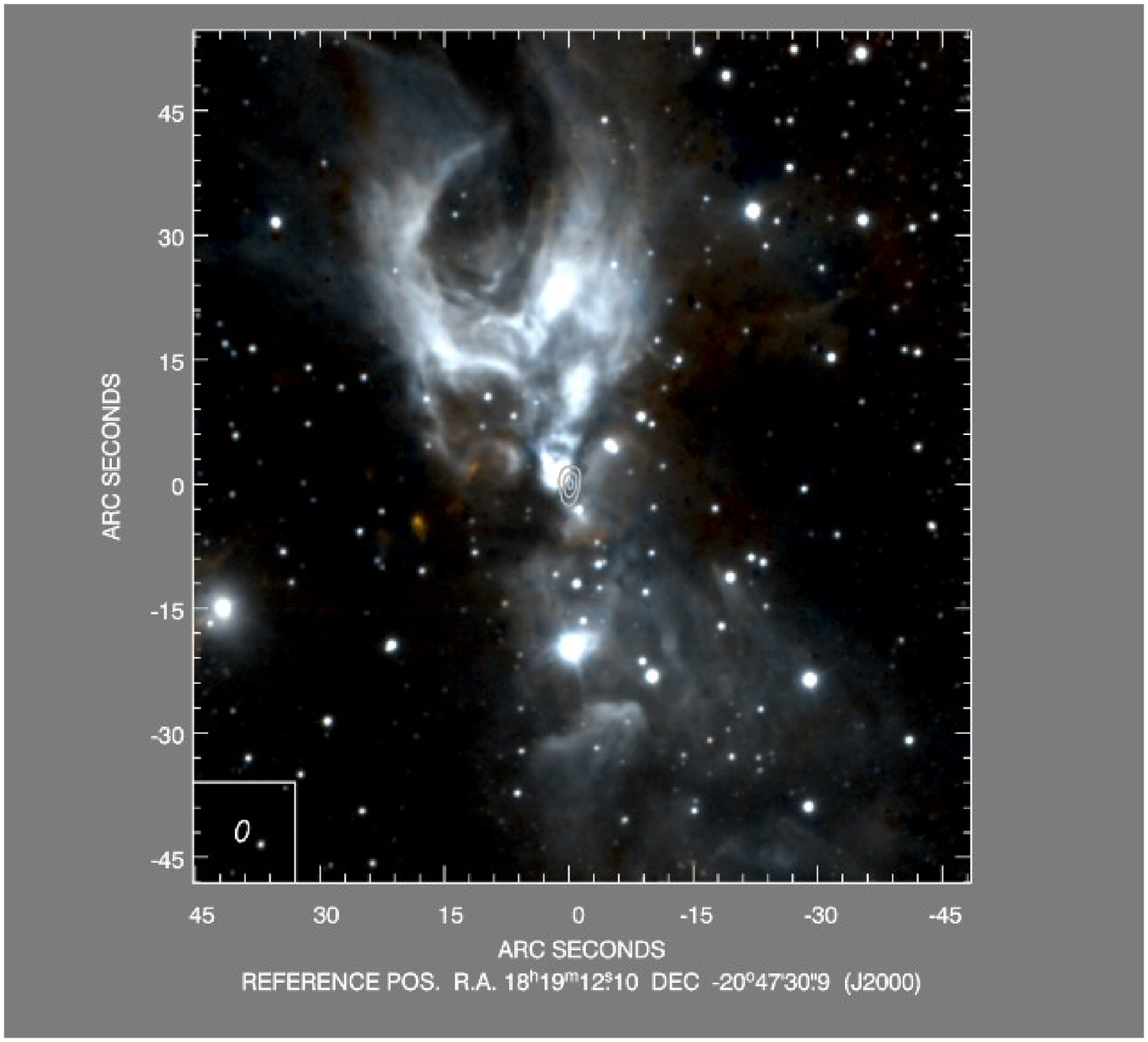}
\vspace*{10cm}
\begin{center}
\parbox{13cm}{\small \bf Figure 4. \rm \,\, The GGD 27 reflection nebula, colour--coded with the H$_2$ emission 
(2.12 $\mu$m) in red and the Br$\,\gamma$ emission (2.17 $\mu$m) in blue. The 
overlaid contours denote the sole emission peak at 7 mm, revealed by our VLA 
D-array data which is coincident with the jet--driving source. It is clearly offset 
from the neighbouring NIR feature IRS2. }
\end{center} 
\vspace*{1cm} 
\includegraphics{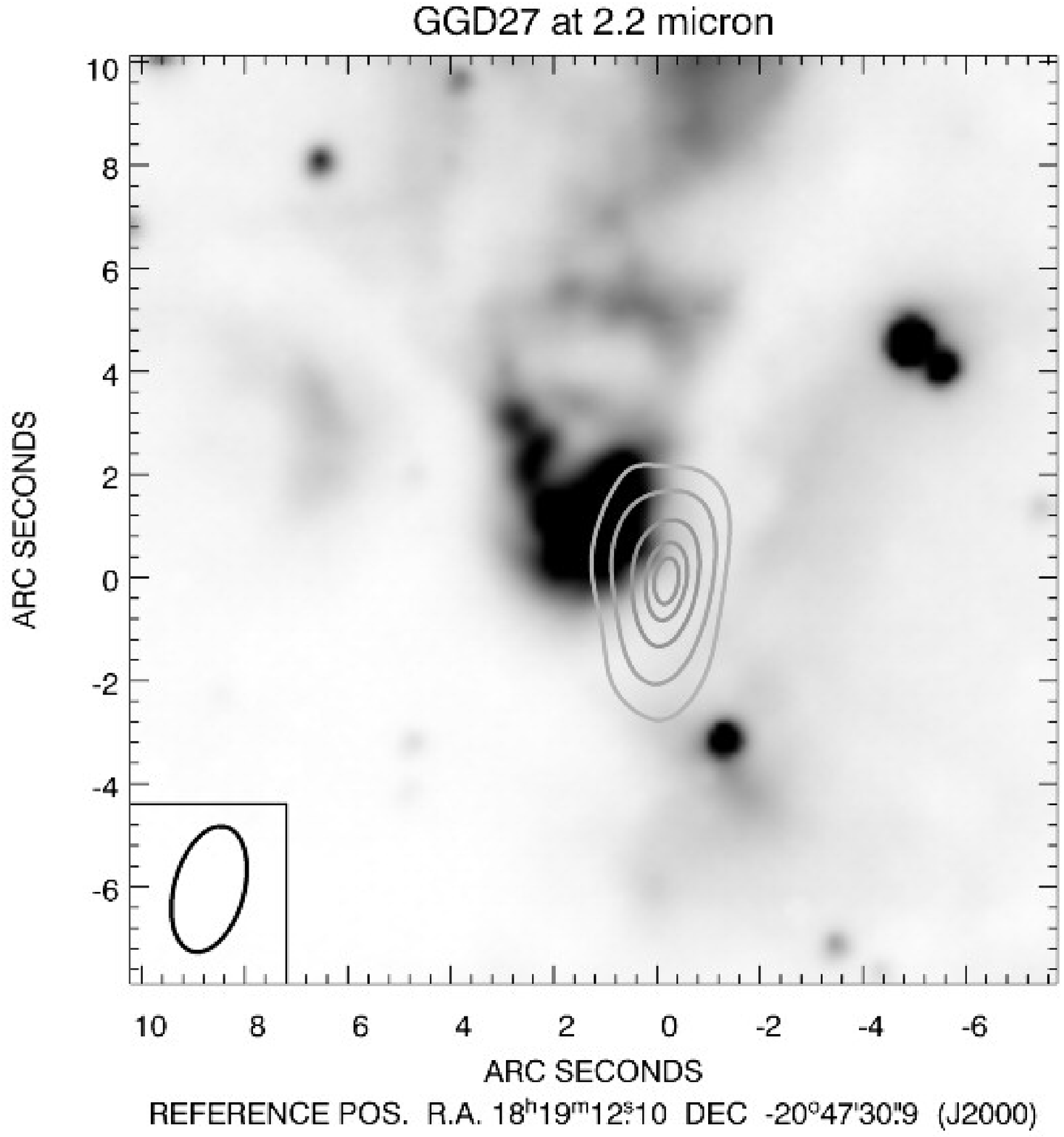}
\includegraphics{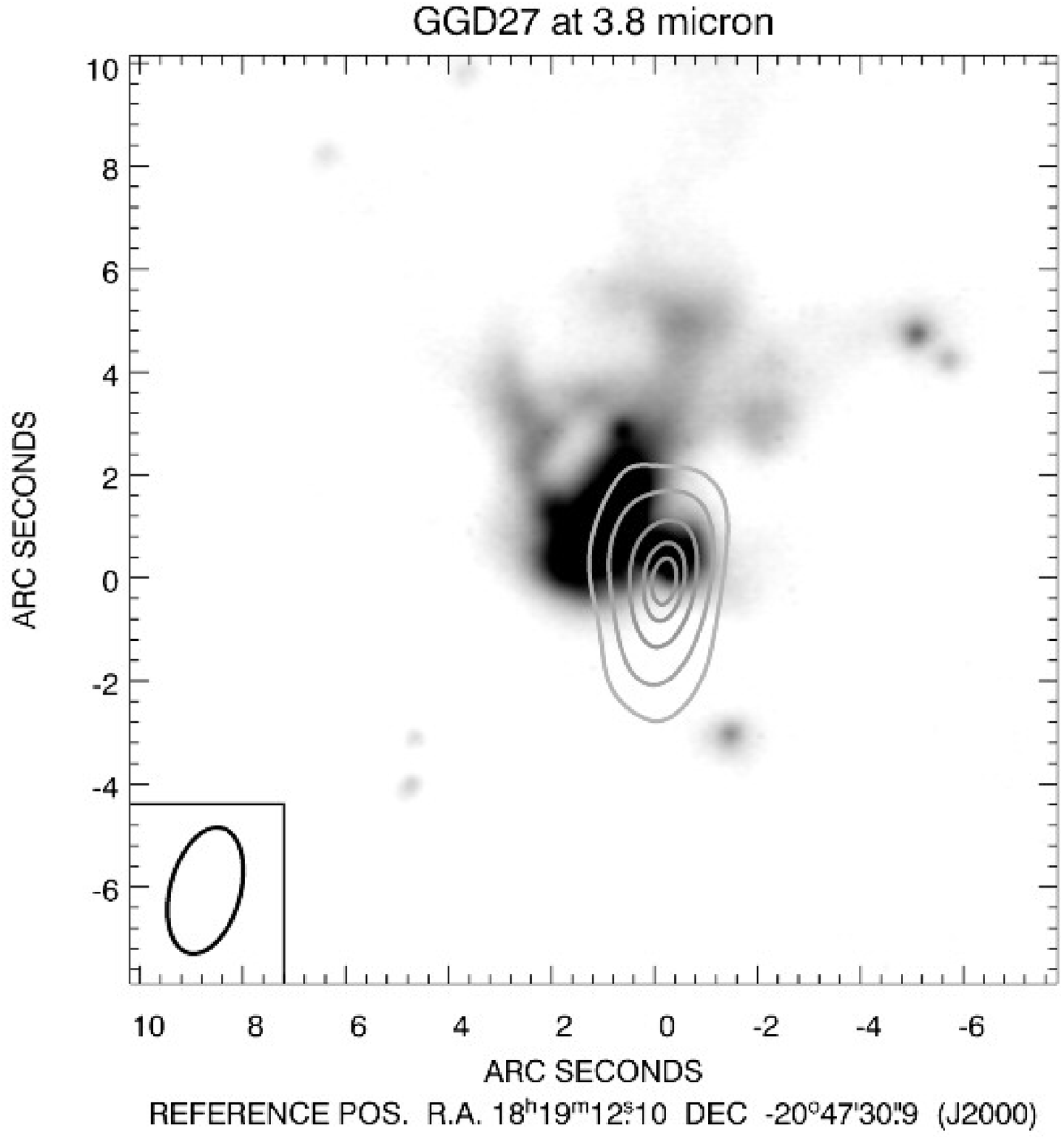}
\includegraphics{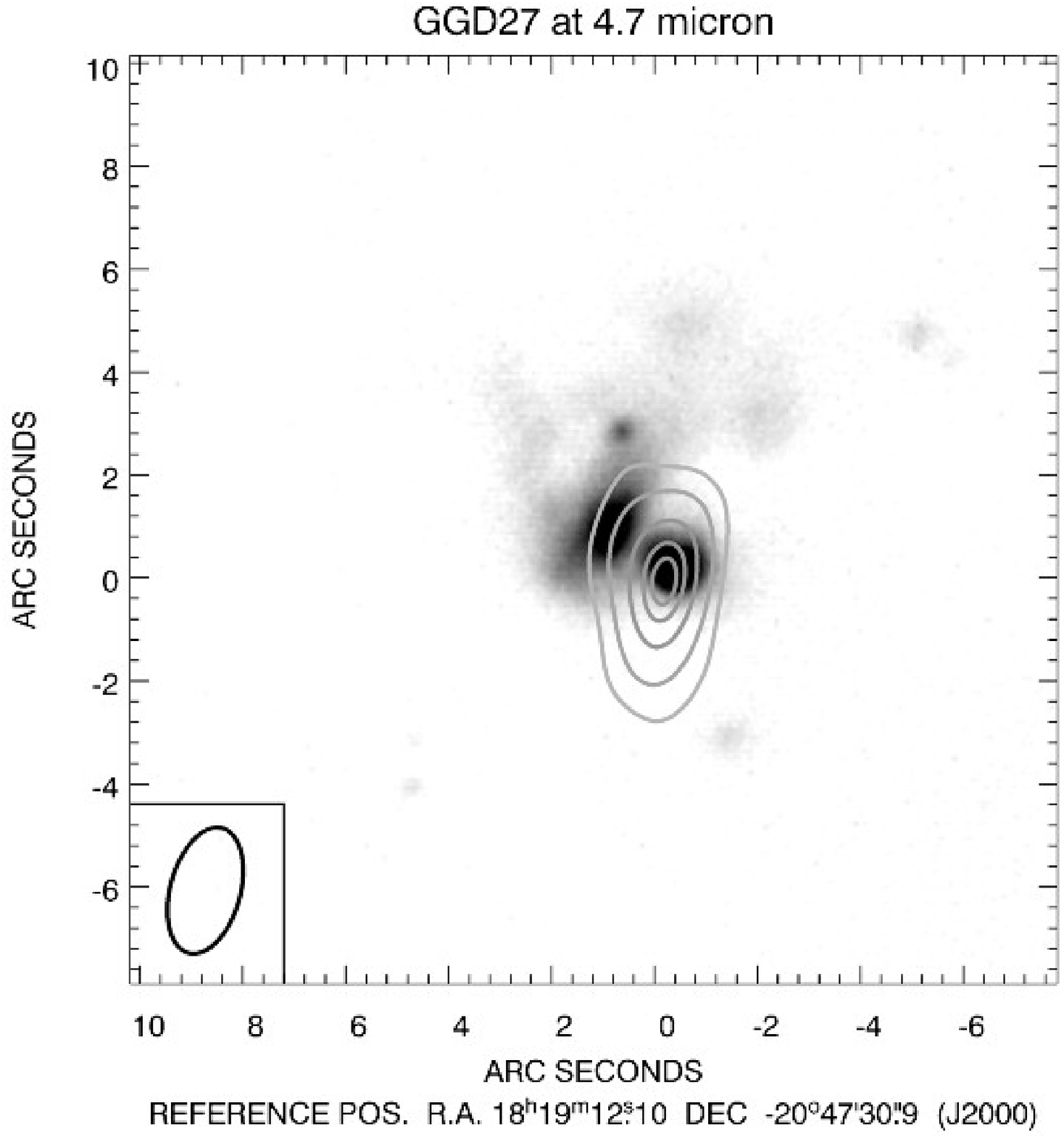}
\vspace*{5.5cm}
\begin{center}
\parbox{13cm}{\small \bf Figure 5. \rm \,\, The centre of the star--forming region GGD27 around IRS2. While still 
hidden at 2.2 micron, the actual power source of the region finally becomes visible 
in the thermal infrared. Again, the contours trace the 7-mm emission which turns out 
to be in good positional agreement with the embedded infrared source.}
\end{center}

\newpage


\section{Conclusions}

As a result of our ISAAC observations, several southern high--mass star--forming
regions could be investigated in greater detail than before. The large degree of
complexity has been revealed. We could show that the actual interesting sources 
often remain invisible in the near--infrared due to their deep embedding in the dense 
gas and dust configurations of the natal molecular clouds. Therefore, imaging in
the thermal infrared turned out to be extremely useful to trace these seclusive
objects. Hence, multi--wavelength data sets were pivotal to get the whole picture.
Accurate astrometry could be applied which has a great impact on the interpretation
of the data. One of the lessons learned is that the relation between infrared and
radio emission is sometimes not as clear--cut as it might seem. Thus, careful and 
detailed investigations like the presented ISAAC campaign are crucial for the deeper 
understanding of the mechanisms of massive star formation. However, the ISAAC results 
represent only an intermediate step. Still higher resolution is required to study all 
the details of massive star formation -- a task for powerful Adaptive Optics systems on 
telescopes of the 8-m class. \\

\acknowledgements H.\,L. and B.\,S. were supported by the German {\em Deut\-sche 
Forschungsgemeinschaft, DFG}, project number STE 605/17-2. P.\,H. acknowledges partial 
support from the Research Corporation grant No CC 4996, as well as from the NSF grant 
AST-0098524. We are indebted to Esteban Araya for the reduction the VLA 7-mm map of GGD27. 
See his poster contribution about the high--mass star--forming region
G31.41+0.31, also on this web page.

\end{document}